\documentclass[pre,twocolumn,aps,eqsecnum]{revtex4-1}
\usepackage{epsfig}

\begin{document}

\title{Shear-Transformation-Zone Theory of Linear Glassy Dynamics}

\author{Eran Bouchbinder}
\affiliation{Department of Chemical Physics, Weizmann Institute of Science, Rehovot 76100, Israel}

\author{J.S. Langer}
\affiliation{Department of Physics, University of California, Santa Barbara, CA  93106-9530}

\date{\today}

\begin{abstract}
We present a linearized shear-transformation-zone (STZ) theory of glassy dynamics in which the internal STZ transition rates are characterized by a broad distribution of activation barriers.  For slowly aging or fully aged systems, the main features of the barrier-height distribution are determined by the effective temperature and other near-equilibrium properties of the configurational degrees of freedom.  Our theory accounts for the wide range of relaxation rates observed  in both structural glasses and soft glassy materials such as colloidal suspensions.  We find that the frequency dependent loss modulus is not just a superposition of Maxwell modes.  Rather, it exhibits an $\alpha$ peak that rises near the viscous relaxation rate and, for nearly jammed, glassy systems, extends to much higher frequencies in accord with experimental observations. We also use this theory to compute strain recovery following a period of large, persistent deformation and then abrupt unloading.  We find that strain recovery is determined in part by the initial barrier-height distribution, but that true structural aging also occurs during this process and determines the system's response to subsequent perturbations. In particular, we find by comparison with experimental data that the initial deformation produces a highly disordered state with a large population of low activation barriers, and that this state relaxes quickly toward one in which the distribution is dominated by the high barriers predicted by the near-equilibrium analysis. The nonequilibrium dynamics of the barrier-height distribution is the most important of the issues raised and left unresolved in this paper.
\end{abstract}
\maketitle

\section{Introduction}
\label{intro}

A growing body of experimental evidence indicates that glassy materials exhibit broad spectra of linear relaxation modes. Most definitively, the frequency dependent loss modulus, $G''(\omega)$, exhibits a broad peak that has a maximum near the viscous relaxation rate and extends over two or more decades of higher frequencies. This behavior has been seen in structural and metallic glasses \cite{GAUTHIER-00,GAUTHIER-04,PELLETIER-06}, and also in soft glassy materials such as colloidal suspensions \cite{93HR, 94MMSZ, 95MW, 95MWa, PURNOMO-07, SIEBENBURGER-09, 10EPDRM, 10CWJCY}. Additional evidence for a broad relaxation spectrum emerges from strain recovery experiments, in which materials are unloaded abruptly after shear deformation \cite{PURNOMO-07, 97OCKB, BELYAVSKY-98}.

Our intention in this paper is to develop a theory that quantitatively accounts for these measurements, extending the results announced in \cite{BL-11}. To date, two of the most successful descriptions of these phenomena have been the soft glassy rheology (SGR) theory of Sollich, Cates, et al. \cite{SOLLICH-97,SOLLICH-98}, and the attempt by Cates and coworkers to use mode-coupling theory (MCT) for similar purposes \cite{BRADER-CATES-08,BRADER-CATES-09}. Here, we base our analysis on the shear-transformation-zone (STZ) theory of amorphous plasticity \cite{FL-98,BLP-07-II,JSL-08,FL-10}. We contend that the STZ theory provides a more suitable starting point than either SGR or MCT. Ultimately, however, our goal is to find a description of linear glassy dynamics that combines the strongest elements of all three of these theories.

In contrast to SGR, which postulates a trap-like mechanism, the STZ theory is based directly on a model of the localized shear transformations that are observed in colloidal suspensions, bubble rafts, and molecular dynamics simulations, and that  presumably occur in molecular glasses as well.  The STZ theory also describes the nonequilibrium flow of energy and entropy in a way that has not yet been achieved in SGR.  On the other hand, the distribution of trapping energies that is an essential component of SGR has, until now, been missing in the STZ theory.  Incorporating this feature of SGR into the STZ theory is a main theme of the present paper.

In contrast to MCT, which starts with a fluidlike many-body Hamiltonian and makes decoupling approximations, the STZ theory is based on a solidlike picture of an amorphous system. This is an intrinsically non-perturbative strategy.  Among other phenomena, this strategy allows the STZ theory to predict a sharp dynamic transition between jammed and flowing states -- a transition that seems difficult to achieve in any perturbation-theoretic analysis that starts with liquidlike variables -- but which is accomplished at least in part in \cite{BRADER-CATES-08,BRADER-CATES-09}.  The success of the STZ theory comes at the expense of a seemingly irreducible degree of phenomenology. Without a precise basis in many-body physics, the STZ theory (like SGR) must postulate a set of relevant internal dynamic variables and use whatever constraints are available -- symmetry principles, the laws of thermodynamics, physical insight, and agreement with experiment -- to deduce equations of motion for those variables.

The STZ theory is based on two fundamental assumptions.  The first of these is that the degrees of freedom of a glassy material can be separated into two subsystems -- the slow configurational degrees of freedom, i.e. the inherent structures \cite{STILLINGER-WEBER-82,STILLINGER-88}, and the fast kinetic-vibrational degrees of freedom which, in the case of colloids, include the motions of the fluid in which the particles are suspended \cite{BLI-09,BLII-09,BLIII-09,BL-Kovacs-10}.  When external perturbations drive the material away from equilibrium, the  effective temperature of the configurational subsystem may depart from the temperature of the kinetic-vibrational subsystem.  The latter temperature generally is the same as the temperature of the thermal reservoir.

The second fundamental premise of the theory is that irreversible shear deformations occur only at rare, localized, two-state, flow defects.  These are the  STZ's which, by definition, belong to the configurational subsystem. In flowing states, the STZ's appear and disappear as the driven system makes transitions between its inherent structures. In jammed states, the STZ's are configurationally frozen, but they still are able to make transitions between their internal states in response to ordinary thermal fluctuations and applied stresses. The two-state nature of the STZ's is chiefly responsible for the dynamic transition between jammed and flowing states at large driving stresses.  As will be seen here in the linear regime, this two-state dynamics -- plus a fundamentally different kinematic starting assumption -- produces a formula for the frequency-dependent modulus $G(\omega)$ that is not the same as that which emerges from SGR.

Earlier versions of STZ theory have been based on the assumption that only one characteristic kind of STZ is needed in order to understand stress-strain relations, shear banding instabilities, and the like.  Thus, our STZ's have had only one statistically significant formation energy, and only a single energy barrier to be crossed during transitions between internal orientational states.  Here we follow \cite{BL-11} and change the latter assumption in order to understand the dynamic measurements; that is, we need to assume that the STZ's occur with a distribution of different internal barrier heights, in much the same way that was deduced experimentally by Argon and Kuo in 1980 \cite{ARGON-KUO-80}.

In the sections that follow, we propose a simple form for an extended, multi-barrier STZ theory, following our earlier report \cite{BL-11}. We restrict our attention to the limit of small external stresses, because that is the limit in which the response of the system is most strongly limited by activation barriers. Accordingly, this small-stress, linear-response limit is where we find the clearest experimental evidence for the multi-barrier picture.  We outline the extended STZ theory in Sec. \ref{STZ-EOM}. In Sec. \ref{oscillatory} we provide a more detailed analysis and additional support for the results announced in \cite{BL-11}, where the frequency dependent modulus $G(\omega)$ was calculated and compared to the data of \cite{GAUTHIER-04,SIEBENBURGER-09}, who performed oscillatory experiments on a wide variety of both hard structural glasses and soft colloidal suspensions.

Finally, in Sec. \ref{recovery}, we further extend the analysis of \cite{BL-11} by looking at the evidence for broadly distributed barrier heights that emerges from strain-recovery measurements and subsequent probes of partially recovered, i.e. partially aged, specimens.  We note that the paper by Belyavsky et al. \cite{BELYAVSKY-98}, dealing with strain recovery in metallic glasses,  contains a clear description of two-state shear transformation zones that predates our own \cite{FL-98}. We focus here on experiments by Purnomo et al  \cite{PURNOMO-07} on colloidal suspensions, in which the strain was measured following a period of large, persistent deformation and then abrupt unloading.  We find that strain recovery is determined in part by the initial barrier-height distribution, but that true structural aging also occurs during this process and strongly determines the subsequent linear responses. In particular, we find that the initial deformation produces a highly disordered state with a large population of low activation barriers, and that this state equilibrates quickly toward one in which the  distribution is dominated by high barriers.

A recurring theme thoughout Secs. \ref{STZ-EOM} - \ref{recovery} is the question of what determines the barrier-height distribution and how this distribution changes during aging. To discuss this issue quantitatively, we start by considering fully aged or slowly aging systems in which these distributions approach states of quasi-equilibrium determined by the current values of the effective temperature and other relevant parameters.  We argue that these distributions are limited by the condition that STZ transition rates cannot be slower than the rates of spontaneous, thermally activated, configurational rearrangements in glassy materials. We then make a phenomenological guess about the form of the barrier-height distribution for younger systems at higher effective temperatures, such as those that occur in the strain-recovery experiments.  In this way, we are able to make some progress in understanding the physics of these phenomena; but we have not yet developed an equation of motion for the STZ barrier-height distribution analogous to the one that serves as a starting assumption in SGR.

Our main conclusions are summarized in Sec.\ref{conclusions}.

\section{Extended STZ Theory}
\label{STZ-EOM}

\subsection{STZ Equations of Motion}

As in almost all earlier presentations, it is easiest and physically most transparent to assume that the STZ's are oriented only in the $\pm$ directions relative to the shear stress $s$.  We lose no generality by doing this; the tensorial generalizations of the equations are obvious at the end of the analysis, but are not needed for present purposes.

In order to describe colloidal suspensions as well as molecular glasses, however, we must recognize that the total stress $s$ acting on the system is the sum of partial stresses  associated with the configurational and kinetic-vibrational subsystems. (See  \cite{BLII-09}.) For the suspensions, the kinetic-vibrational partial stress is the viscous stress generated by hydrodynamic forces. Therefore,
\begin{equation}
s = s_C + \eta_K \ast \dot\gamma.
\end{equation}
where $s_C$ is the configurational partial stress, and $\dot\gamma$ is the total shear rate, common to both the configurational and kinetic-vibrational subsystems.  The notation $\eta_K \ast$ means that the viscosity $\eta_K$ is a time-retarded  integral operator that becomes a function of frequency after Fourier transformation.

The new feature that was introduced in \cite{BL-11} is that the STZ's are characterized by internal barrier heights, say, $\Delta$.  Let the number of $\pm$ STZ's with given $\Delta$ be $N_{\pm}(\Delta)$, and let the total number of (coarse-grained) molecular sites be $N$.  In the limit of small applied stresses, the master equation for $N_{\pm}(\Delta)$ is
\begin{eqnarray}
\label{Ndot}
\nonumber
\tau_0\,\dot N_{\pm}(\Delta) &=& R(\pm s_C,\Delta)\,N_{\mp}(\Delta) - R(\mp s_C,\Delta)\,N_{\pm}(\Delta)\cr\\& +& \rho(\theta)\,\left[{N_{eq}(\Delta)\over 2} - N_{\pm}(\Delta)\right].
\end{eqnarray}
Here, $R(\pm s_C,\Delta))/\tau_0$ is the rate per STZ for transitions between $\pm$ orientations; it is the origin of the $\Delta$ dependence in the theory. Only the partial stress $s_C$ appears as an argument of $R(\pm s_C,\Delta))$ because the STZ's are configurational defects. $\tau_0$ is a fundamental time scale, for example, a vibration period for molecular glasses or a Brownian diffusion time for colloidal suspensions.

The last two terms on the right-hand side of Eq.(\ref{Ndot}) are the rates at which STZ's are created and annihilated by spontaneous thermal fluctuations.  We are making a detailed-balance approximation in which $N_{eq}(\Delta)/2$ is the value approached by $N_{\pm}(\Delta)$ in steady-state equilibrium.  $\rho(\theta)$ is the thermal noise strength, and $\theta = k_B\,T$ is the bath temperature in units of energy.  Mechanically generated noise, which is a prominent ingredient of earlier analyses \cite{JSL-08,FL-10}, is second order in the applied stress and therefore is negligible in this linear theory.

As usual \cite{FL-10}, define the internal state variables:
\begin{eqnarray}
\label{Lambda-m-def}
\nonumber
\Lambda(\Delta) &=& {N_+(\Delta)+N_-(\Delta)\over N};\cr\\m(\Delta)&=&{N_+(\Delta)-N_-(\Delta)\over N_+(\Delta) +N_-(\Delta)}.
\end{eqnarray}
According to Eq.(\ref{Ndot}), the equations of motion for these variables are
\begin{equation}
\label{Lambdadot}
\tau_0\,\dot \Lambda(\Delta) = \rho(\theta)\,\Bigl[\Lambda_{eq}(\Delta) - \Lambda(\Delta)\Bigr],
\end{equation}
where $\Lambda_{eq}(\Delta) = N_{eq}(\Delta)/ N$;
\begin{eqnarray}
\label{mdot}
\nonumber
\tau_0\,\dot m(\Delta) &=& 2\,{\cal C}(s_C,\Delta)\Bigl[{\cal T}(s_C,\Delta) - m(\Delta)\Bigr] \cr \\ &-& \rho(\theta)\,m(\Delta) -{\tau_0\,\dot\Lambda(\Delta)\over \Lambda(\Delta)}\,m(\Delta);
\end{eqnarray}
and
\begin{eqnarray}
\nonumber
{\cal C}(s_C,\Delta)&=& {1\over 2}\,\Bigl[R(s_C,\Delta) + R(-s_C,\Delta)\Bigr];\cr\\{\cal T}(s_C,\Delta)&=& {R(s_C,\Delta) - R(-s_C,\Delta)\over R(s_C,\Delta) + R(-s_C,\Delta)}.
\end{eqnarray}
The total rate of plastic deformation is a superposition of terms of the form
\begin{eqnarray}
\label{DplastDelta}
\nonumber
&&\tau_0\,D^{pl}(\Delta) = {v_0\over V}\,\Bigl[R(s_C,\Delta)\,N_-(\Delta)-R(-s_C,\Delta)\,N_+(\Delta)\Bigr]\cr\\ &&= \epsilon_0\,\Lambda(\Delta)\,{\cal C}(s_C,\Delta)\Bigl[{\cal T}(s_C,\Delta) - m(\Delta)\Bigr],
\end{eqnarray}
where $V$ is the volume of the system, and $v_0$ is a molecular volume that sets the size of the plastic strain increment induced by an STZ transition. We expect  $\epsilon_0 \equiv N\,v_0/V$ to be a number of the order of unity.

\subsection{Glassy Dynamics}

Most -- but not all -- of the linear responses of interest here are associated with STZ transitions between their internal states. The exceptions are the last two terms on the right-hand side of Eq.(\ref{mdot}), which derive from the annihilation and creation terms in Eq.(\ref{Ndot}), and therefore do describe configurational changes.  As discussed in earlier papers, $\Lambda_{eq}=\exp\,(-\,e_Z/\chi)$, where $\chi$ is the effective disorder temperature in energy units, and $e_Z$ is an STZ formation energy that, at this stage of the discussion, may be a $\Delta$-dependent quantity. Thus, in Eq.(\ref{Lambdadot}), the rate at which STZ's are spontaneously created (and annihilated) by thermal fluctuations is proportional to the factor
\begin{equation}
\label{chi-activation}
\dot\Lambda_{{\rm creation}}(\Delta) \propto{\rho(\theta)\over \tau_0} \,e^{-\,e_Z(\Delta)/\chi},
\end{equation}
which has the form of an activation rate, with $\chi$ playing the role of the temperature, and $\rho(\theta)/\tau_0$ being the attempt frequency.

Equation (\ref{chi-activation}) embodies some of the deepest issues in glass physics.  It expresses our contention that the glass transition is intrinsically a dynamic phenomenon, rather than a thermodynamic phase transition. Thus, the formation energy $e_Z$ has the magnitude of an ordinary interaction energy; and the Boltzmann factor $\exp\,(-\,e_Z/\chi)$ determines the probability of finding a fluctuation of energy $e_Z$ in the configurational subsystem once it has come to equilibrium at temperature $\chi$.  On the other hand, the dynamic prefactor $\rho(\theta)$ is a super-Arrhenius function of the temperature, associated with the fact that the configurational rearrangements needed to form an STZ-like defect involve many-body fluctuations that become increasingly complex and unlikely as the temperature decreases. (See \cite{JSL-XCHAINS-06} for one hypothetical picture of this mechanism.)  Accordingly, $\rho(\theta)$ is approximately equal to unity at temperatures well above the nominal glass temperature $\theta_g$, but it becomes very small at lower temperatures.  The system undergoes a dynamically sharp glass transition at some temperature, say, $\theta_0 < \theta_g$, if -- as in the Vogel-Fulcher-Tamann formula -- $\rho(\theta)$ actually vanishes at and below $\theta_0$.

The difference between $\chi$ and $\theta$ is a measure of the extent to which the configurational degrees of freedom have fallen out of equilibrium with the heat bath. In the absence of mechanically generated noise, configurational, i.e. ``structural,'' relaxation is governed by an equation of motion for $\chi$, which we write in the form
\begin{equation}
\label{chidot}
\tau_0\,{\dot\chi\over e_Z} = \kappa\,\rho(\theta)\,e^{-\,e_A/\chi}\,\left(1-{\chi\over \theta}\right).
\end{equation}
Here $\kappa$ is a dimensionless constant, very roughly of the order of unity, and $\exp\,(-\,e_A/\chi)$ is a measure of the population of defects, with formation energy $e_A$, that enable energy transfer from the configurational to the kinetic-vibrational degrees of freedom. The energy $e_A$ should be roughly the same as $e_Z$. Thus, Eq.(\ref{chidot}) describes slow, structural aging during which $\chi$ relaxes toward $\theta$.

With these assumptions, we introduce a normalized distribution over barrier heights, $p(\Delta)$, and use Eq.(\ref{DplastDelta}) to write the total rate of plastic deformation in the form
\begin{equation}
\label{Dplast}
D^{pl}(s_C) = {\epsilon_0\over \tau_0}\langle \Lambda \rangle\int d\Delta\,p(\Delta)\,{\cal C}(s_C,\Delta)\,\left[{\cal T}(s_C,\Delta)-m(\Delta)\right].
\end{equation}
Here, because no small factor of the form $\exp\,(-\,e_Z(\Delta)/\chi)$ appears in the prefactor in Eq.(\ref{Lambdadot}), we have assumed that $\Lambda(\Delta) \cong \Lambda_{eq}(\Delta)= \exp\,(-\,e_Z(\Delta)/\chi)$.
Then, for simplicity, we have assumed that we can bring the latter factor outside of the integral over $\Delta$ in Eq.(\ref{Dplast}), and replace $e_Z(\Delta)$ by a single characteristic formation energy $e_Z$. Thus we have written
\begin{equation}
 \langle \Lambda \rangle \equiv \exp\,(-\,e_Z/\chi).
\end{equation}

We now need to specify the transition rate $R(s_C,\Delta)$.  In earlier work \cite{JSL-08,FL-10}, we wrote this quantity in the form
\begin{equation}
\label{Rdef}
R(s_C,\Delta)= R_0(s_C)\,\exp \left[-\,{{\cal D}(s_C)\over \theta}\right],
\end{equation}
where the function ${\cal D}(s_C)$ was equal to the single barrier height $\Delta$ when $s_C=0$ and became vanishingly small at large, positive $s_C$.  Here, with a range of different $\Delta$'s, we must recognize that the stress needed to drive the system over a barrier of height $\Delta$ must itself be a growing function of $\Delta$.  The simplest such choice is
\begin{equation}
\label{calD}
{\cal D}(s_C) = \Delta\,\exp\left(-\,{v_0\,s_C\over a_0\,\Delta}\right),
\end{equation}
where $v_0$, needed for dimensional reasons, can be taken to be the same molecular volume introduced in Eq.(\ref{DplastDelta}), and $a_0$ is a dimensionless number with which we account for the uncertainties in the other parameters. If the assumptions leading to this equation are correct, then $a_0$ will be roughly equal to unity. The approximation for ${\cal D}(s_C)$ in Eq.(\ref{calD}) is physically reasonable and mathematically well behaved except for small $\Delta$, i.e. for high rates, where it will play no role in the analyses to be described here

The prefactor $R_0(s_C)$ can, in principle, be any symmetric function of $s_C$.  However, since we are considering only very small values of $s_C$, this stress dependence is irrelevant.  On the other hand, because some STZ's are complex objects with many internal degrees of freedom, we anticipate that the attempt frequency $R_0/\tau_0$ depends on both the temperature $\theta$ and the activation barrier $\Delta$.  Therefore, in analogy to the STZ creation-rate formula in Eq.(\ref{chi-activation}), we write $R_0 = \rho_0(\theta,\Delta)$.  Once again, we presume that the activation barriers $\Delta$ are ordinary energies roughly comparable to the formation energies $e_Z$;  but that the rates of thermally assisted passage over these barriers may be substantially decreased by glassy dynamics. The $\Delta$ dependence of $\rho_0$ is necessary. We expect that the transitions over small barriers, near $\Delta = 0$, are simple Arrhenius processes with $\rho_0 \cong 1$.  At the other extreme, when $\Delta$ is large, we expect $\rho_0$ to be small.

Note that the separation of time scales between the configurational and kinetic-vibrational degrees of freedom requires that the internal STZ transitions be no slower than configurational rearrangements; that is,
\begin{equation}
\label{rho0ineq}
\rho_0(\theta,\Delta)\,e^{-\,\Delta/\theta} > \rho(\theta)\,e^{-\,e_Z/\theta}.
\end{equation}
We anticipate that, even if there were a glass transition at a nonzero temperature $\theta_0$, the prefactor  $\rho_0(\theta,\Delta)$ would remain nonzero for $\theta < \theta_0$ although $\rho(\theta)$ would vanish.

These understandings about the rate factors allow us to simplify the linearized STZ equations of motion. For small $s_C$,
\begin{equation}
\label{C}
{\cal C}(s_C,\Delta) \cong {\cal C}(0,\Delta) = \rho_0(\theta,\Delta)\,e^{- \Delta/\theta},
\end{equation}
and
\begin{equation}
{\cal T}(s_C,\Delta) \cong {\cal T}'(0)\,s_C = {v_0\,s_C\over a_0\,\theta}.
\end{equation}
We stress that ${\cal C}(s_C,\Delta)$ in Eq.(\ref{C}) describes only ordinary thermally activated processes.

It is convenient to write
\begin{equation}
\label{nudef}
2\,{\cal C}(0,\Delta)= 2\,\rho_0(\theta,\Delta)\,e^{- \Delta/\theta}\equiv \nu(\Delta),
\end{equation}
and, in most circumstances, to use $\nu$ as the independent variable instead of $\Delta$. The barrier heights $\Delta$ are energies that characterize the configurational subsystem.  They do not, by themselves, carry dynamic information.  In contrast, when we write equations in terms of $\nu$, we are building into them a large amount of information about glassy dynamics.  Thus, the transformation of variables in Eq.(\ref{nudef}) is an important feature of the following analysis.

Define
\begin{equation}
\label{tilde-p}
\tilde p(\nu) = -\,p(\Delta)\,{d\Delta\over d\nu}.
\end{equation}
Then, up to terms linear in $s_C$, Eq.(\ref{Dplast}) becomes
\begin{equation}
\label{Dplast2}
D^{pl}(s_C) = {\epsilon_0\over 2\,\tau_0}\langle \Lambda \rangle\int d\nu\,\tilde p(\nu)\,\nu\,\left[{v_0\,s_C\over a_0\,\theta}-\tilde m(\nu)\right],~~~~~~~
\end{equation}
where $\tilde m(\nu) = m(\Delta)$.  The equation of motion for $\tilde m$, Eq.(\ref{mdot}), becomes
\begin{equation}
\label{mdot2}
\tau_0\,\dot {\tilde m}(\nu) = {v_0\,\nu\over a_0\,\theta}\,s_C - (\nu+ \rho)\,\tilde m(\nu),
\end{equation}
where we have dropped the term proportional to $\dot\Lambda$ on the right-hand side of Eq.(\ref{mdot}).

\section{Oscillatory Response}
\label{oscillatory}

\subsection{Frequency-Dependent Modulus $G(\omega)$}

We start by assuming that we are dealing with systems that are sufficiently well aged that $\langle \Lambda \rangle$ has approached its equilibrium value, and no longer is changing as a function of time at rates comparable to experimental oscillation periods.  We also assume -- nontrivially -- that the total shear rate $\dot\gamma$ is simply the sum of elastic and plastic parts:
\begin{equation}
\label{gammadot}
\dot\gamma = {\dot s_C\over \mu} + D^{pl}(s_C);~~~~s_C = s -\,\eta_K\ast\dot \gamma;
\end{equation}
where $\mu$ is the shear modulus.

Denote Fourier transforms as functions of frequency $\omega$ by $\hat m$, $\hat s$, etc.; and let
\begin{equation}
\eta_K\ast\dot \gamma \to i\,\omega\,\hat \eta_K(\omega)\,\hat\gamma.
\end{equation}
Then,
\begin{equation}
\hat m(\nu) = {v_0\,\nu\over a_0\,\theta}\,\left[{\hat s - i\,\omega\,\hat\eta_K(\omega)\,\hat\gamma\over i\,\omega\,\tau_0 + \nu + \rho}\right].
\end{equation}
Similarly, Eq.(\ref{gammadot}) becomes
\begin{equation}
i\,\omega\,\hat\gamma = \left[{i\,\omega\over \mu}+ {\epsilon_0\,v_0\,\langle\Lambda\rangle\over 2\,a_0\,\theta\,\tau_0} \,J(\omega)\right]\,
\left[\hat s -i\,\omega\,\hat\eta_K(\omega)\hat\gamma\right],
\end{equation}
where
\begin{equation}
\label{Jdef}
J(\omega) = \int d\nu\,\tilde p(\nu)\,\nu\,\left({i\,\omega\,\tau_0 +\rho\over i\,\omega\,\tau_0 +\rho +\nu}\right).
\end{equation}
Solving for $G(\omega) = \hat s/\hat\gamma$, we find
\begin{equation}
\label{G}
G(\omega) = i\,\omega\,\tau_0\,\mu\,\left[{{\cal N}(\omega)\over i\,\omega\,\tau_0 + \bar\Lambda\,J(\omega)}\right];
\end{equation}
where
\begin{equation}
{\cal N}(\omega) = 1 + {i\,\omega\over \mu}\,\hat\eta_K(\omega)  + {\hat\eta_K(\omega)\over \mu\,\tau_0}\,\bar\Lambda\,J(\omega)
\end{equation}
and
\begin{equation}
\label{Lambdabar}
\bar\Lambda = {\epsilon_0\,v_0\,\mu\over 2\,a_0\,\theta}\,\langle\Lambda\rangle\cong {\epsilon_0\,v_0\,\mu\over 2\,a_0\,\theta}\,e^{-\,e_Z/\chi}.
\end{equation}

To take a first look at these formulas, assume that the kinetic viscosity $\eta_K$ is negligible at low frequencies, and use Eq.(\ref{G}) to compute the Newtonian viscosity associated with configurational deformation:
\begin{equation}
\label{etaN}
\eta_N = \lim_{\omega \to 0}\,{G(\omega)\over i\,\omega} = {\mu\,\tau_0\over  \bar\Lambda\,J(0)}.
\end{equation}
It is easy to check that, when the distribution $\tilde p(\nu)$ is sharply peaked at one characteristic value of $\nu$, this formula reduces to the STZ result in Eq.(5.4) of \cite{JSL-08}.

Next, return to the expression for $G(\omega)$ given in Eq.(\ref{G}), and note that this formula cannot naturally be expressed as an average over Maxwell modes as in SGR.  This feature is a result of our kinematic assumption in Eq.(\ref{gammadot}), plus our assumption that the plastic strain rate appearing there is a sum over independent contributions from the different kinds of STZ's.

We can deduce immediately from Eq.(\ref{G}), without yet knowing much about the distribution function $\tilde p(\nu)$, that the low-frequency structure of  $G(\omega)$ occurs approximately in the neighborhood of
\begin{equation}
\label{omega-alpha}
\omega_{\alpha} \sim {\bar\Lambda\,J(0)\over \tau_0}= {\mu\over \eta_N},
\end{equation}
so that $\omega_{\alpha}$ is about the same as the viscous relaxation rate. However, the structure of the $\alpha$ peak depends sensitively on the barrier-height distribution, and the approximation made in Eq.(\ref{omega-alpha}) cannot replace a full evaluation of $G(\omega)$. For example, if $\tilde p(\nu)$ were concentrated in a narrow band of values near, say, $\nu = \nu_0 \gg \rho$, there would be a relatively narrow peak in $G''(\omega)$ at $\omega = \omega_{\alpha}$ as given by Eq.(\ref{omega-alpha}), but $G(\omega)$ would have other structure at and above $\omega = \nu_0/\tau_0$.  As will be seen, an anomalously broad $\alpha$ peak requires that $\tilde p(\nu)$ have an appreciable part of its support starting well below $\nu = \omega_{\alpha}\,\tau_0$.

\subsection{Barrier-Height Distribution}

At this point, we must begin to discuss the barrier-height distribution $p(\Delta)$. Our starting assumption is that $p(\Delta)$ is a near-equilibrium feature of the configurational subsystem. By this, we mean that it is determined by configurational variables, especially the effective temperature, which themselves may be changing in time while the system as a whole moves toward true thermodynamic equilibrium.  Thus, we start by considering systems that are fully aged or are aging slowly, and later will see what happens when this assumption fails.

Since $\Delta$ is measured {\it downward} from some reference energy, it seems natural to postulate, at least for a range of values of $\Delta$, that $p(\Delta)$ is an equilibrium distribution of the form
\begin{equation}
\label{pDeltalow}
p(\Delta) \propto e^{+\Delta/\tilde\Delta},
\end{equation}
where $\tilde\Delta$ is an energy that will be proportional to $\chi$ in simple circumstances. For thermosensitive colloidal suspensions, however, we expect that $\tilde\Delta$ will depend on the volume fraction, which itself is a strongly varying function of the bath temperature $\theta$.  A relation of the form $\tilde\Delta \propto \chi$ would mean that the distribution of energy barriers becomes broader as $\chi$ increases, consistent with the fact that $\chi$ is a measure of disorder as well as energy. In the limit of small $\Delta$ and large $\nu$, where $\rho_0 = 1$, the factor $d\,\Delta/d\,\nu$ in Eq.(\ref{tilde-p}) is equal to  $ -\,\theta/\nu$, and
\begin{equation}
\label{pnuhigh}
\tilde p(\nu) \approx{\tilde A\over \nu^{1+\zeta}},~~~~\zeta = {\theta\over\tilde\Delta},
\end{equation}
where $\tilde A$ is a normalization factor.

In the opposite limit of large $\Delta$ and small $\nu$, integrability of $p(\Delta)$ requires that the distribution be cut off for, say, $\Delta > \Delta^*$, $\nu < \nu^*$. To estimate $\nu^*$, remember that the inequality in Eq.(\ref{rho0ineq}) tells us that the STZ transition rates must be faster than the rates at which the STZ's are created and annihilated by thermal fluctuations.  It obviously makes no sense to talk about STZ transition rates that are slower than the rates at which the STZ's themselves are appearing and disappearing.  We therefore propose that $\nu^*$ be the value of $\nu$ at which this inequality breaks down, i.e.
\begin{equation}
\label{nu*}
\nu^* = 2\,\rho_0(\theta,\Delta^*)\,e^{-\,\Delta^*/\theta} = 2\,\rho(\theta)\,e^{-\,e_Z/\theta}.
\end{equation}

Importantly, we do not need to know anything at all about $\rho_0$ in order to determine $\nu^*$ from the last expression on the right-hand side of Eq.(\ref{nu*}).  This relation predicts extremely small values of $\nu^*$, which are confirmed by experiments. The presence of $\rho_0$ in the intermediate expression simply assures us that we do not have to invoke values of $\Delta^*$ that are unphysically larger than $e_Z$ in order to justify  small values of $\nu^*$.

The cutoff for $\Delta > \Delta^*$, $\nu < \nu^*$ cannot be infinitely sharp; therefore, in this region, we propose to write
\begin{equation}
\label{pDeltahigh}
p(\Delta)\propto e^{-\,\Delta/\tilde\Delta_1},
\end{equation}
and, equivalently, in the limit of small $\nu$,
\begin{equation}
\tilde p(\nu) \approx {\tilde A\over \nu^{1 - \zeta_1}}, ~~~~ \zeta_1 = {\theta\over \tilde\Delta_1}.
\end{equation}
Here, $\tilde\Delta_1$ is an as-yet undetermined energy scale, and we have assumed (questionably) that the $\Delta$ dependence of $\rho_0$ is unimportant in this limit. In general, we expect that $\zeta_1 \sim 1$. A much larger $\zeta_1$ -- i.e. a much sharper cutoff -- would require the configurational energy scale $\tilde\Delta_1$ to be smaller than $\theta$, which seems unphysical. On the other hand, substantially smaller values of $\zeta_1$ would be inconsistent with our rationale for the choice of $\nu^*$ in Eq.(\ref{nu*}).  (In Sec.\ref{recovery}, we will invoke small values of $\zeta_1$ for highly disordered systems far from configurational equilibrium, where the preceding rationale for $\nu^*$ is no longer valid.) The resulting structure of the function $p(\Delta)$, on both sides of its peak at $\Delta^*$, is remarkably similar to the distributions deduced from creep measurements by Argon and Kuo \cite{ARGON-KUO-80}.

Throughout the rest of this paper, we combine the high- and low-$\nu$ approximations by writing
\begin{equation}
\label{pnu}
\tilde p(\nu) \cong {\tilde A\over \nu\,\left[(\nu/\nu^*)^{\zeta} + (\nu^*/\nu)^{\zeta_1}\right]}.
\end{equation}
This is an overly simple, three-parameter representation of the barrier-height distribution, with two exponents $\zeta$ and $\zeta_1$ determining the large-$\nu$ and small-$\nu$ limits respectively, and a single crossover value of $\nu^*$.

The parameter $\zeta$ controls the high-frequency behavior of $G(\omega)$.  For $\omega\,\tau_0 \gg \rho$,
\begin{equation}
G(\omega)\approx {\mu\over 1 + \bar\Lambda\,J(\omega)/i\,\omega\,\tau_0},
\end{equation}
and
\begin{equation}
\label{J/omega}
{J(\omega)\over i\,\omega\,\tau_0} \approx \int_0^2 {\nu\,\tilde p(\nu)\,d\nu\over \nu + i\,\omega\,\tau_0} \propto \int_{\nu^*}^2 {\nu^{-\,\zeta}\,d\nu \over \nu + i\,\omega\,\tau_0}.
\end{equation}
A scaling analysis, using $\nu^*\ll 2$, then tells us that
\begin{equation}
\label{J/omega-2}
{J(\omega)\over i\,\omega\,\tau_0} \propto {C(\zeta)\over (\omega\,\tau_0)^{\zeta}},
\end{equation}
where $C(\zeta)$ is a complex constant.  Therefore, $G''(\omega) \sim (\omega\,\tau_0)^{-\zeta}$ in the limit of large $\omega$.

Suppose, for the moment,  that $\tilde\Delta$ is simply proportional to $\chi$, say, $\tilde\Delta = \chi/b$, where $b$ is a system-dependent parameter.  Suppose, further, that the system is truly in equilibrium, so that $\chi = \theta$.  Then Eq.(\ref{pnuhigh}) implies that $\zeta = b$.  This parameter ``$b$'', or one nearly equivalent, has been computed from first principles in MCT \cite{FUCHS-10}.  Perhaps this is a place where we can find a direct relation between the STZ and MCT theories.

\begin{figure}
\centering \epsfig{width=.52\textwidth,file=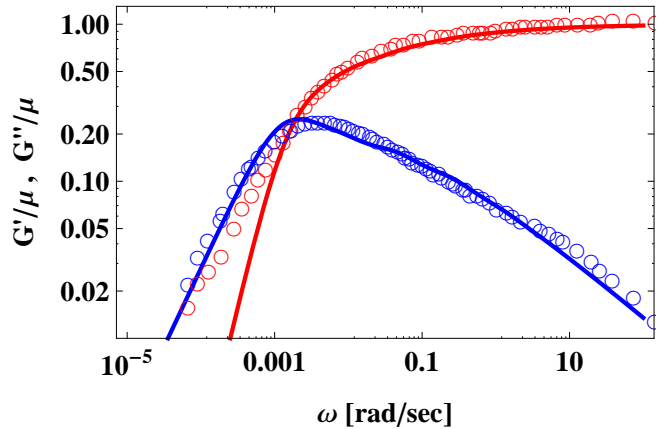} \caption{(Color online) Experimental data for Vitreloy 4 and theoretical comparisons for the storage modulus $G'(\omega)$ (red) and the loss modulus $G''(\omega)$ (blue).  The data points are taken directly from  Fig.2 of Gauthier et al. \cite{GAUTHIER-04}. } \label{G-BMG}
\end{figure}

\subsection{Structural and Metallic Glasses}

In comparing these theoretical results with experimental data, we look first at the oscillatory response of structural and metallic glasses, for which $\tau_0$ is of the order of picoseconds, $\omega\,\tau_0 \ll 1$, and $\eta_K$ is negligible.  The interesting behavior occurs at temperatures near or slightly above the glass temperature.

Our principal sources of information about the oscillatory responses of structural and metallic glasses are the papers by Gauthier et al., in particular \cite{GAUTHIER-04}.   These authors show that the functions $G(\omega)$, for a wide variety of noncrystalline materials at their glass temperatures, have very similar behaviors. Specifically, the loss modulus $G''(\omega)$ has a broad peak at $\omega_{\alpha}$  and drops off at high frequencies like $\omega^{-\zeta}$ as predicted in Eq.(\ref{J/omega-2}). For metallic glasses, Gauthier et al. find $\zeta \cong 0.4$.

In Fig.\ref{G-BMG}, we show  $G'(\omega)/\mu$ and $G''(\omega)/\mu$ as predicted by Eq.(\ref{G}), along with data from Fig.2 of \cite{GAUTHIER-04}, for the metallic glass Vitreloy 4 at its glass temperature $T_g$.  In estimating the theoretical parameters, we have used $T_g \cong 600\,K$, $\tau_0 \cong 2 \times 10^{-12}$ sec., and $\mu \cong 50\,{\rm GPa}$. To make approximations for the other parameters in Eq.(\ref{G}), we note that, if the volume $v_0$ is of the order of a few cubic nanometers, then the ratio $v_0\,\mu /\theta_g$, and thus the prefactor in Eq.(\ref{Lambdabar}), is approximately $10^4$.  Then, to estimate $\langle\Lambda(\theta_g)\rangle \sim \exp\,(-\,e_Z/\theta_g)$, we assume that $\theta_g$ is the same as the steady-state value of the effective temperature, usually denoted by $\chi_0$, for systems driven persistently at shear rates much slower than $\tau_0^{-1}$.  The ratio $\chi_0/e_Z$ may be a universal quantity; it usually turns out to be in the range $0.1 - 0.2$. We therefore estimate $\theta_g/e_Z \sim 0.15$, implying that $\langle\Lambda(\theta_g)\rangle \sim 10^{-3}$, and therefore that $\bar\Lambda \sim 10$. As a result, Eq.(\ref{nu*}) tells us that $\nu^* \cong 10^{-3}\,\rho(\theta_g)$.

The theoretical curves in Fig.\ref{G-BMG} have been computed using  $\zeta_1 = 1$,  $\rho(\theta_g)/\tau_0= 1.25 \times 10^{-2}\,\,{\rm sec.}^{-1}$, and $\bar\Lambda = 25$.  In effect, we have set $\epsilon_0 / a_0 \sim 2.5$ in Eq.(\ref{Lambdabar}), which is well within our theoretical uncertainty.  Note that these parameters imply that $\nu^*\sim 10^{-17}$, which, as predicted, is extremely small compared, for example, to its upper limit at $\nu = 2$. So far as we can tell from numerical exploration, this small value of $\nu^*$ is sharply determined by the experimental data. Because $\tilde p(\nu)$ is varying very rapidly near $\nu = \nu^*$, changing $\nu^*$ by even a factor of 2 ruins the fit to the data; and the curves become qualitatively wrong if $\nu^*$ is changed by an order of magnitude in either direction. Our theoretical curve for $G''(\omega)$ in Fig.1 is somewhat less smooth than the experimental curve, which may be a result of our overly simple form for $\tilde p(\nu)$ in Eq.(\ref{pnu}).  The other discrepancy is that the experimental  storage modulus $G'(\omega)$ is approximately linear in $\omega$ at low frequencies, instead of being proportional to $\omega^2$ as predicted by our theory as well as by Maxwell models.

\begin{figure}[here]
\centering \epsfig{width=.52\textwidth,file=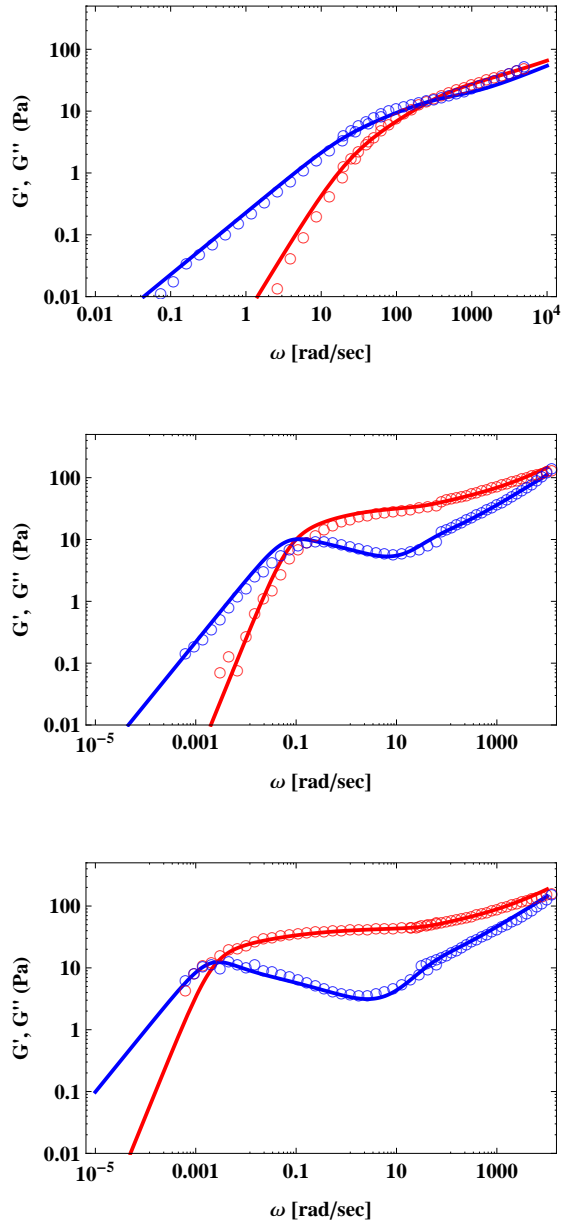} \caption{(Color online) Experimental data and theoretical comparisons for the storage modulus $G'(\omega)$ (red) and the loss modulus $G''(\omega)$ (blue), for three different suspensions of thermosensitive particles, as reported in \cite{SIEBENBURGER-09}. The values of the parameters are listed in the text. } \label{G-Colloid}
\end{figure}

\subsection{Colloidal Suspensions}

The rheology of soft, colloidal suspensions differs from that of structural and metallic glasses in at least two important respects.  First, in colloidal systems, the approach to jamming near a glass transition is controlled more sensitively by the volume fraction than by the temperature. Second, the microscopically short molecular vibration period in structural glasses is replaced in colloids by the very much longer time scale for Brownian motion of the particles.  As a result, the high-frequency cutoff at $\nu = 2$ is probed in rheological experiments, and the kinetic viscosity $\hat\eta_K(\omega)$ is relevant at accessibly high values of $\omega$.  When samples are prepared by subjecting them to strong shear stresses, their shear rates become comparable to $\tau_0^{-1}$.  Therefore, they become highly disordered and their effective temperatures $\chi$ can become arbitrarily large \cite{JSL-MANNING-TEFF-07}.

In order to use Eq.(\ref{G}) to evaluate $G(\omega)$ for colloidal suspensions, we need an expression for the frequency-dependent kinetic viscosity $\hat\eta_K(\omega)$.  Here, we follow Lionberger and Russel \cite{LIONBERGER-RUSSEL-94}, who show that $G'(\omega) \approx \omega^{1/2}$ in the limit of very large $\omega$.  Their analysis is based on the idea that, in order to satisfy the no-flow boundary condition at the surface of a colloidal particle, the surrounding fluid must form a diffusive boundary layer whose thickness scales as $\omega^{-1/2}$.  The viscous force on this particle therefore decreases like $\hat\eta_K(\omega)\sim \omega^{-1/2}$, and the resulting stress is $i\,\omega\,\hat\eta_K(\omega)\sim \omega^{1/2}$.  We regularize this expression at low frequencies by assuming that
\begin{equation}
\label{etaK}
\hat\eta_K(\omega)= {\mu\,\tau_K\over (c + i\,\omega\,\tau_K)^{1/2}}
\end{equation}
where $c$ is a dimensionless constant, and $\tau_K$ is a viscous time scale.

In Fig.\ref{G-Colloid}, we show three examples of how the STZ theory developed here is capable of reproducing the experimental results of Siebenburger et al. \cite{SIEBENBURGER-09}. These authors explored a range of effective volume fractions $\phi_{e\!f\!f}$ and a wide range of frequencies $\omega$ (as well as steady shear rates not discussed here) by using suspensions of thermosensitive particles (polystyrene cores with attached networks of thermosensitive isopropylacrylamide molecules). They have used the effective size of the particles and the solvent viscosity to estimate that the Brownian time scale is $\tau_0 \sim .003$ secs. \cite{SIEBENBURGER-FUCHS-10}.

The sequence of increasing volume fractions, from the top panel to the bottom in Fig.\ref{G-Colloid}, is $\phi_{e\!f\!f} = 0.518,\,0.600,\,{\rm and}\, 0.626$. The theoretical parameters, deduced by fitting the data and listed in the same order, are $\rho = 0.04,\,3\times 10^{-4},\, 10^{-5}$; $\nu^* = 0.001,\,10^{-3}\rho,\,10^{-3}\rho$; $\zeta = 1.0,\,0.5,\,0.38$; $\bar\Lambda = 200,\,40,\,17$; $\mu = 12,\,35,\,45\,{\rm Pa}$; and $\tau_K = 0.004,\,0.002,\,0.002$ sec..  In all cases, $\zeta_1 = 1$ and $c=0.1$.

The trends are interesting.  The sequence of examples starts in the top panel of Fig.\ref{G-Colloid} with a system whose relatively small volume fraction puts it well away from the glass transition.  It is effectively a liquid; $\rho = 0.04$ means that there is relatively little super-Arrhenius suppression of the structural relaxation rate.  As the systems become more glassy in the middle and bottom panels, $\rho$ decreases rapidly.  $\zeta$ also decreases, as if the temperature $\theta$ in Eq.(\ref{pnuhigh}) were decreasing; but it is the increasing volume fraction that must be causing the energy scale $\tilde\Delta$ to increase.  At the same time, the shear modulus $\mu$ increases and the viscous time scale $\tau_K$ decreases slightly as the systems become stiffer.  $\bar\Lambda$ is very large for the liquidlike example in the top panel, implying that the STZ density is large in this system. On the other hand, the values of $\bar\Lambda$, and the relation between $\nu^*$ and $\rho$ for the two nearly glassy cases, are almost the same as the analogous estimates for bulk metallic glass in the preceding subsection -- despite the fact that the underlying time scales for these systems differ by nine orders of magnitude.

\section{Strain Recovery and Aging}
\label{recovery}

\subsection{Strain Recovery: Basic Aspects}

The success of the STZ theory in accounting for a wide range of oscillatory measurements gives us some confidence that our fundamental concepts are correct.  We turn now to a class of experiments whose interpretation is not nearly so simple.

Although strain-recovery experiments technically probe the linear response theory outlined in the preceding sections, they are qualitatively different from the oscillatory experiments because the mode of sample preparation is necessarily nonlinear.  In order to induce a strain whose recovery can be observed, these systems are deformed at stresses well above the yield stress. As a result, we cannot use near-equilibrium concepts to estimate the initial barrier-height distribution as we did for the oscillatory analysis. Nor can we assume that the initial value of the effective temperature $\chi$ is close enough to a steady-state value that we can neglect its time dependence.

\begin{figure}
\centering \epsfig{width=.52\textwidth,file=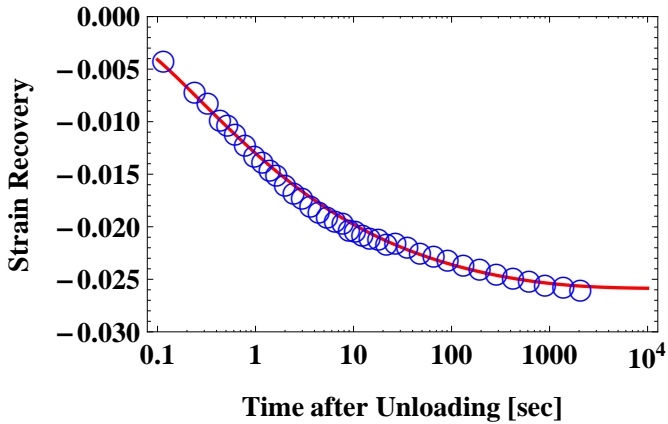} \caption{(Color online) Strain recovery as a function of time.  The blue data points are taken directly from Fig.4b of \cite{PURNOMO-07}. The strain is measured from its value after unloading and after an initial, almost instantaneous, elastic relaxation. The red theoretical curve has been computed using the parameter values discussed in the text. } \label{Strain-Recovery}
\end{figure}

Strain recovery measurements have been carried out by Belyavsky et al. \cite{BELYAVSKY-98} for metallic glasses, and by Purnomo et al. \cite{PURNOMO-07} for colloidal suspensions. We focus on the more recent colloidal experiments described in \cite{PURNOMO-07}; and we further specialize to results for the thermosensitive poly-N-isopropylacrylamide microgel suspension that Purnomo et al. call ``P-1''.  A representative data set taken from Fig.4b in  \cite{PURNOMO-07} is shown here in  Fig.\ref{Strain-Recovery}.

In both metallic glass and colloidal cases, the systems first were deformed at high stress, and then unloaded quickly. After an initial, almost instantaneous, elastic relaxation, the plastic strain decreased slowly at zero applied stress, as seen here in Fig.\ref{Strain-Recovery}.  We compute this plastic relaxation by setting $s_C = 0$ in  Eq.(\ref{mdot2}), obtaining
\begin{equation}
\label{mdot3}
\tau_0\,\dot{\tilde m}(\nu) = - (\nu +\rho)\,\tilde m(\nu);~~\tilde m(\nu) = \tilde m_0\,e^{-\,(\nu+\rho)\,t/\tau_0},~~~~
\end{equation}
where, for simplicity, we assume that the initial values of all the $\tilde m(\nu)$ are equal to $\tilde m_0$. Then, using Eqs.(\ref{Dplast2}) and (\ref{gammadot}), in both cases with $s_C = 0$ and no kinetic viscosity, we have
\begin{eqnarray}
\label{gamma-t1}
\nonumber
&&\dot\gamma = - {\epsilon_0\over 2\,\tau_0}\,\langle\Lambda\rangle\,\int_0^2 d\nu\,\tilde p(\nu)\,\nu\,\tilde m(\nu,t)\cr \\&& =-\,{\epsilon_0\,\tilde m_0\over 2\,\tau_0}\,\langle\Lambda\rangle\int_0^2 d\nu\,\tilde p(\nu)\,\nu\,e^{-\,(\nu+\rho)\,t/\tau_0}.~~~~~~
\end{eqnarray}
A first, rough comparison between these formulas and the experimental strain-recovery data tells us that we are seeing an entirely different distribution $\tilde p(\nu)$ than the one that we used to interpret the oscillatory data in Sec.\ref{oscillatory}.

One clue about what might be happening is contained in a paper by Rodney and Schuh \cite{RODNEY-SCHUH-09}.  These authors used Monte Carlo simulations to track the distributions of thermally activated events in a binary, two dimensional, glassy material. When their system was allowed to age at sub-yield stresses, they found decreasing numbers of increasingly high activation barriers. In contrast, when they persistently deformed their  system at high stresses, they found larger numbers of lower barriers. Karmakar et al. \cite{10KLP} have reported similar observations about barriers seen in athermal quasi-static simulations.

In the language used here in Sec.\ref{oscillatory}, with the notation introduced in Eq.(\ref{pnu}) for the distribution $\tilde p(\nu)$, the first case discussed by Rodney and Schuh  \cite{RODNEY-SCHUH-09} resembles the small-$\nu^*$ distribution at small $\chi$ that we used to describe the oscillatory responses of well aged systems. In their second case, they drove their system to large $\chi$, i.e. to high energy and high disorder, where they saw large numbers of low activation barriers. This case corresponds to a large value of $\nu^*$, with the weight of the distribution in the range $\nu < \nu^*$, and with $\tilde p(\nu) \sim \nu^{-1 + \zeta_1}$. The approximation for $\nu^*$ in Eq.(\ref{nu*}) is not relevant in this far-from-equilibrium situation, where the barrier heights near $\nu^*$ are too small to be limited by thermally induced configurational fluctuations.

Suppose, for the moment, that we can ignore the time dependence of $\langle\Lambda\rangle$ in Eq.(\ref {gamma-t1}).  Then
\begin{eqnarray}
\label{gamma-t2}
\nonumber
&&\gamma(t) = - \,{\epsilon_0\,\tilde m_0\over 2}\,\langle\Lambda\rangle\,\int_0^2 d\nu\,{\nu\,\tilde p(\nu)\over \nu + \rho}\,\left[1 - e^{- \,(\nu + \rho)\,t/\tau_0}\right]\cr \\&&\equiv -\,\gamma_{\infty} + {\epsilon_0\,\tilde m_0\over 2}\,\langle\Lambda\rangle\int_0^2 d\nu\,{\nu\,\tilde p(\nu)\over \nu + \rho}\,e^{-\,(\nu+\rho)\,t/\tau_0}.~~~~
\end{eqnarray}
Thus, after unloading to zero stress, the strain ultimately relaxes by an amount $\gamma_{\infty}$, and does so at a rate that is no slower than $\exp\,(-\,\rho\,t/\tau_0)$. For large values of $t/\tau_0$, the integral over $\nu$ in Eq.(\ref{gamma-t2}) is dominated by the part of $\tilde p(\nu)$ for $\nu < \nu^*$, and therefore:
\begin{eqnarray}
\label{gamma-t3}
\nonumber
\gamma(t)&\approx& -\gamma_{\infty} + {\epsilon_0\,\tilde m_0\over 2}\,\langle\Lambda\rangle\,\tilde A_c\,e^{-\,\rho\,t/\tau_0}\int_0^{\tau_0/t} d\nu\,\nu^{\zeta_1-1}\cr \\ &=& - \,\gamma_{\infty} +{{\rm const.}\over (t/\tau_0)^{\zeta_1}}\,e^{-\,\rho\,t/\tau_0}.
\end{eqnarray}
If $\zeta_1$ is small, that is, if the distribution $\tilde p(\nu)$ is nearly flat down  to small values of $\nu$, then strain recovery will appear to be logarithmic out to times of the order of $\tau_0/\rho$, which is the behavior described by Belyavsky et al.\cite{BELYAVSKY-98}.

\subsection{Stress-Step Experiments and Structural Aging}

\begin{figure}
\centering \epsfig{width=.52\textwidth,file=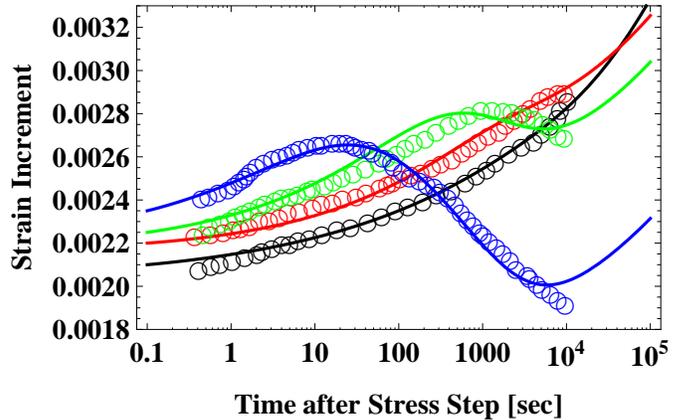} \caption{(Color online) Strain increments induced by small stress steps at waiting times of $t_w = 30,\, 600,\, 3,000\, {\rm and}\, 10,000$ secs, indicated respectively, from top to bottom on the left-hand side, by blue, green, red, and black data points and theoretical curves . The data points are taken directly from Fig.5a of  \cite{PURNOMO-07}. The parameters used in computing the theoretical curves are discussed in the text.} \label{Step-Stress}
\end{figure}

The strain-recovery data for the colloidal suspensions of Purnomo et al.  \cite{PURNOMO-07} can be fit reasonably well by using Eq.(\ref{gamma-t3}) with values of $\zeta_1$ in the range $0.2\,-\,0.3$.  With only this information, we might have concluded that these authors were seeing just the $\tilde m$ relaxation described by Eq.(\ref{mdot3}), and not true structural aging in which STZ's are created and annihilated.  But these authors went further in important ways. After preparing their samples by rapid straining and abrupt unloading, they allowed the strain to relax at zero stress for various waiting times $t_w$, and then measured the response to a small step in the stress whose magnitude $\delta s_C$ was much less than the yield stress. Their results are shown here in Fig.\ref{Step-Stress}.

At least two aspects of these results can be understood only in terms of true structural aging.  First, the initial, instantaneous, elastic increase in the strain, following the stress step, decreased with increasing aging time, indicating that the shear modulus $\mu$ was larger in the older samples.  The shear modulus must be a decreasing function of $\chi$ (just as it is a decreasing function of the ordinary temperature $\theta$). Thus, we conclude that $\chi$ -- an intrinsically structural variable -- decreased during aging. Second, Purnomo et al. found that, during the initial responses to the stress steps, the functions $\gamma(t)$ for different waiting times could be scaled approximately onto each other if they were plotted as functions of $t/t_w$, where $t = 0$ is the time at which  the stress step was applied.  Both of these features can be seen here (roughly) in Fig.\ref{Step-Stress}.

To interpet these behaviors, consider the equation of motion for $\gamma$ for times $t > 0$. If Eq.(\ref{mdot3}) were accurate throughout the aging process -- which we will argue is not exactly the case -- then
\begin{equation}
\tilde m(\nu,t=0) = \tilde m_0\,e^{-(\nu + \rho)t_w/\tau_0}.
\end{equation}
We insert this formula into Eq.(\ref{Dplast2}), and find that the subsequent rate of deformation is
\begin{eqnarray}
\label{dotgamma-step}
\nonumber
\dot\gamma &=& {\langle\Lambda(t+t_w)\rangle\over \tau_0}\,\int_0^2 d\nu\,\tilde p(\nu)\,\nu\,\Bigl\{-\, \bar m \,e^{-\,(\nu + \rho)(t+t_w)/\tau_0}\cr\\&+& \sigma\,\Bigl[{\rho\over \nu+\rho} + {\nu\over \nu + \rho}\,e^{-\,(\nu + \rho)t/\tau_0}\Bigr]\Bigr\},
\end{eqnarray}
where
\begin{equation}
\bar m = {\epsilon_0\over 2}\,\tilde m_0,~~~~\sigma = {\epsilon_0\,v_0\,\delta s_C\over 2\,a_0\,\theta}.
\end{equation}
The first term inside the brackets, proportional to $\bar m$, is the continuation of the original strain-recovery formula following the waiting time $t_w$. Of the two terms proportional to the dimensionless stress step $\sigma$, the first is the viscous response, consistent with Eq.(\ref{etaN}), which is nonzero so long as $\rho$ is nonzero.  The last term tells us that, when $\rho$ is small, this system initially behaves like a conventional yield-stress material; that is, for small stresses, it fairly quickly becomes jammed with a step-wise, fixed increase in the plastic deformation. In contrast, the viscous strain proportional to $\rho$ continues to grow linearly in time, and thus can become arbitrarily large.

A crucial point is that each of the terms in Eq.(\ref{dotgamma-step}) is proportional to the aging factor $\langle \Lambda (t+t_w) \rangle = \exp\,[-\,e_Z/\chi(t+t_w)]$.  For comparisons with experimental data, we must compute this function exactly by solving Eq.(\ref{chidot}) numerically; but it is useful to look first at an analytic approximation.  For simplicity, assume that $e_A = e_Z$ , and also assume that we are interested primarily in situations where $\chi$, while decaying, is still much larger than $\theta$.  Then, in the limit $K\,t\gg 1$, Eq.(\ref{chidot}) has the asymptotic solution
\begin{equation}
\label{chi-t-approx}
e^{-\,e_Z/\chi(t)} \approx {1\over K\,t\,\ln(K\,t)},~~~~ K = {\kappa\,e_Z\,\rho\over \tau_0\,\theta}.
\end{equation}
Therefore, apart from the logarithmic correction in Eq.(\ref{chi-t-approx}), all the terms on the right-hand side of Eq.(\ref{dotgamma-step}) contain the factor $(t+t_w)^{-1} \approx t_w^{-1}$ for $t \ll t_w$, which is roughly consistent with the observed behavior.

\subsection{Preliminary Comparisons between Theory and Experiment}

We turn, finally, to quantitative comparisons between our theory and the experimental data.  We will argue that, although we are able to achieve fairly good agreement with the experiments, the remaining discrepancies indicate that the theory is missing an important ingredient.  Specifically, we still need an equation of motion to describe the time dependence of the barrier-height distribution $\tilde p(\nu)$ during structural aging.

We have used Eq.(\ref{chidot}), along with Eqs.(\ref{gamma-t1}) and (\ref{dotgamma-step}) for strain recovery and stress-step response respectively, to interpret the experimental data shown in Figs. \ref{Strain-Recovery} and \ref{Step-Stress}.  Figure \ref{Strain-Recovery} shows data for strain recovery taken directly from Fig. 4b  of \cite{PURNOMO-07}. Figure \ref{Step-Stress} shows data taken directly from Fig. 5a of  \cite{PURNOMO-07} for responses to stress steps applied after waiting times of $30,\,600,\,3,000,\,{\rm and}\, 10,000$ sec..

Consider first the comparison between theory and experiment in Fig.\ref{Strain-Recovery}.  Using values for the particle radius and solvent viscosity given in  \cite{PURNOMO-07}, we have estimated that $\tau_0 \cong 0.1$ sec..  In solving Eq.(\ref{chidot}) for the ratio $\chi(t)/e_Z$, we have set $e_A = e_Z$ and, assuming that the system is near its glass temperature, have used $\theta/e_Z = 0.15$ as discussed in Sec. \ref{oscillatory} C.  We have evaluated the rate factor $K$ defined in Eq.(\ref{chi-t-approx}) by assuming that the shear modulus $\mu(\chi)$ is a linear function of the form $\mu_0 - \mu_1\,\chi$, and then fitting this relation, with $\chi = \chi(t_w)$, to the values of the initial elastic increments seen in Fig. 5a of  \cite{PURNOMO-07}.  The result is that $K \sim 10^{-3}\,{\rm sec.}^{-1}$, which already is interesting  because it implies that the rate at which $\chi$ relaxes to $\theta$ in Eq.(\ref{chidot}) is comparable to the rate of strain recovery seen in Fig. \ref{Strain-Recovery}.  With this value of $K$, and with the preceding estimates of $\tau_0$ and $\theta/e_Z$, we guess that $\kappa \sim 0.1$ and thus have chosen $\rho\sim 10^{-6}$.  We also -- by necessity -- have assumed that the volume fraction remained fixed throughout these experiments.

The theoretical curve shown in Fig.\ref{Strain-Recovery} has been computed using $\rho = 3 \times 10^{-6}$, $K = 10^{-3}$, $\bar m= \epsilon_0\,\tilde m_0/2 = 0.157$, $\zeta = 0.4$, $\zeta_1 = 0.32$, $\chi(0)/e_Z = 1.0$, and $\nu^* = 0.5$.  With these choices of parameters, the agreement between theory and experiment seems good; but, in fact, it is quite problematic. Most obviously, $\nu^* = 0.5$ is vastly different from the value $\nu^* \sim 10^{-3} \rho \sim 10^{-9}$ that we expect to find in nearly equilibrated systems. A related problem is that we have assumed that the barrier-height distribution $\tilde p(\nu)$ remains fixed during this process; but $\tilde p(\nu)$ is a structural property, and must change during structural aging of the kind implied by decreasing values of $\chi(t)$.  Aging somehow causes the system to make a transformation from its initial, highly disordered state, with a barrier-height distribution characterized by large values of $\chi$ and $\nu^*$, to more ordered states with substantially smaller values of those parameters, especially the latter.  As yet, we have no theory of how this strongly nonequilibrium process occurs.  We do not know, for example, whether $\nu^*$ might be a near-equilibrium function of the instantaneous value of $\chi$, or whether it relaxes toward its near-equilibrium value on some dynamically interesting time scale as happens in \cite{BLP-07-II}, or whether any such weakly nonequilibrium mechanism is operative.

Turn now to the stress-step data in Fig.\ref{Step-Stress}.  For well aged systems such as those for $t_w = 3,000\, {\rm and}\, 10,000$ secs., it seems safe to assume that the configurational subsystems have settled into states that are much the same as those discussed in Secs. \ref{oscillatory} C and D, with $\nu^* \sim 10^{-3} \rho$.  Because of this qualitative change in $\tilde p(\nu)$, there is no reason to expect that the system's orientational memory, carried by the parameter $\bar m$ in Eq.(\ref{dotgamma-step}), retains its original value; thus we use it as a fitting parameter, and find that it does undergo modest changes during aging.

We have used the following procedure to determine the theoretical parameters for the four cases shown in Fig.\ref{Step-Stress}. As we did for the strain-recovery curve in Fig.\ref{Strain-Recovery}, we have computed each of these curves separately by choosing constant values of $\rho$, $K$, $\nu^*$, $\zeta$, $\zeta_1$, and $\bar m$, rather than trying to guess how these parameters might depend on the time.  We have kept $\rho = 3 \times 10^{-6}$ and $K = 10^{-3}$ from the strain-recovery analysis. Despite the fact that $\nu^*$ changes dramatically during the early stages of aging,  we have chosen $\nu^* = 10^{-3}\,\rho$ for all four cases. This relation has a theoretical rationale, and the data does not compel us to choose otherwise -- not even for $t_w = 30$ sec.. We looked first at the most fully aged system, with $t_w = 10,000\,\,{\rm sec.}$.  Here, as argued in Sec.\ref{oscillatory} D, we have assumed that $\zeta_1 = 1$.  We then have fit the data by choosing $\zeta = 0.21$, $\bar m = 0.09$, and $\sigma = 0.07$.  This value of $\sigma$ must be common to all four curves in Fig.\ref{Step-Stress}.

For the three other waiting times, the values of the parameters are: for $t_w = 3,000\,\,{\rm sec.}, \,\zeta = 0.29,\,\zeta_1 = 1,\,\bar m = 0.092$; for $t_w = 600\,\,{\rm sec.}, \,\zeta = 0.28,\,\zeta_1=0.5,\,\bar m = 0.125$; for $t_w = 30\,\,{\rm sec.}, \,\zeta = 0.23,\,\zeta_1= 0.2,\,\bar m = 0.130$.   As expected, the values of $\bar m$ decrease with increasing $t_w$, but they differ from the value ($0.157$) deduced from the strain-recovery data by less than $50\%$.  The numerical results for $t_w = 30\,\,{\rm and}\,\, 600\,\,{\rm sec.}$ are sensitive to the value of $\zeta_1$, which seems to increase with increasing $t_w$.  We see no pattern in the values of $\zeta$, which may be artifacts of our parameterization of $\tilde p(\nu)$ or of our parameter-fitting procedure. They might also be explained by changes in the volume fraction during aging.  There are other experimental uncertainties.  The data shown in Fig.5a of  \cite{PURNOMO-07} is very noisy.  As a result, our estimates of the initial elastic strains in Fig.\ref{Step-Stress} are uncertain; and we do not know how seriously to take other discrepancies between theory and experiment at later times.

One systematic theoretical error is that, for the two cases with shorter aging times, the rising viscous strain is too large at the latest times observed. We have emphasized this behavior in Fig.\ref{Step-Stress} by extending the time axis an extra decade beyond the data.  The formula for the viscosity in Eq.(\ref{etaN}) depends on $\tilde p(\nu)$. In these cases, $\tilde p(\nu)$ has been chosen primarily to fit the earlier stages of the measurements; thus, it is not surprising that the viscosity is wrong at the later stages.

The most interesting feature of these results is the huge change in $\nu^*$ that apparently occurs in the first ten seconds or so after unloading. It appears that the population of STZ's with low activation barriers in the strongly disordered inital state disappears quickly -- much more quickly than can be described by near-equilibrium $\chi$ dynamics -- so that these STZ's are not available to respond to the only slightly delayed step in the stress at $t_w = 30$ sec..  This observation suggests that, even if we could deduce an equation of motion for $\nu^*$, we might not capture essential features of the configurational dynamics in this regime. We then must ask, if  $\tilde p(\nu)$ is such a strong function of aging time, why are we able to describe strain recovery with a time-independent $\tilde p(\nu)$?  Perhaps the behavior seen at long times in Fig.\ref{Strain-Recovery} is sensitive only to the population of slow STZ's, and perhaps this population remains unchanged by aging during the times of interest.  But such an explanation does not yet provide a basis for developing a predictive theory of these phenomena under far-from-equilibrium conditions.

\section{Summary and Conclusions}
\label{conclusions}

The analyses described here lead us to a number of conclusions and assertions.  Some of these, such as (3) and (5), are shared by SGR and MCT, but emerge here in different ways.  Others seem to differ more strongly from earlier formulations.\\

\noindent (1) Linear response in glassy systems is determined by thermally activated processes.  Long time scales can be explained, in analogy to structural relaxation in glasses, by super-Arrhenius reduction of attempt frequencies as opposed to anomalously high activation barriers. \\

\noindent (2) STZ dynamics and kinematics, with a nontrivial barrier-height distribution, produces a non-Maxwellian form of the frequency-dependent modulus $G(\omega)$.\\

\noindent (3) The loss modulus $G''(\omega)$ automatically has an $\alpha$ peak in the neighborhood of the viscous relaxation rate, independent of any specific distribution of barrier heights.\\

\noindent (4) For fully aged or slowly aging systems, the principal features of the STZ barrier-height distribution $\tilde p(\nu)$ can be determined by assuming that the system is nearly in equilibrium with the current values of various configurational variables, most importantly, with the effective temperature $\chi$. However, a detailed, first-principles relation between $\tilde p(\nu)$ and $\chi$ remains undetermined by the analysis presented here, even for equilibrium situations.\\

\noindent (5) Our near-equilibrium analysis of the barrier-height distribution in well aged glassy materials implies that {\it any} such material exhibits an anomalously broad $\alpha$ peak near its glass temperature.  Metallic and structural glasses, and colloidal suspensions, all share remarkably similar properties in this regard, despite the enormous differences in their underlying time scales and internal dynamics.\\

\noindent (6) Some, but not all, of the dynamics of slow processes such as strain recovery can be attributed to internal orientational relaxation, rather than to true structural aging.\\

\noindent (7) True structural aging is characterized by rearrangements of configurational degrees of freedom. In particular, the barrier-height distribution must change during structural aging or persistent deformation.  At present, we still need an equation of motion for $\tilde p(\nu)$, perhaps analogous to the SGR equation of motion for the distribution over barrier heights and local strains. The nonequilibrium dynamics of the barrier-height distribution is the most important of the issues raised and left unresolved in this paper. We hope to return to it in the near future.\\

The analyses presented here have not touched on several other topics that often are included in rheological discussions.  For example, we have not discussed stress relaxation in fixed-strain experiments.  Among the subtle issues that arise in this connection is that the Fourier transform of the stress relaxation function is not necessarily the same as the function $G(\omega)$ defined here. The difference is that our systems contain relevant internal variables such as $\tilde m(\nu)$ and $\chi$, which relax on the same time scales as the stress, so that conventional theories with history-independent relaxation kernels are inapplicable.

We also have not discussed nonlinear plasticity, because that topic has been the main focus of recent STZ theories.  For example, the curves of stress versus strain rate shown by Siebenburger et al \cite{SIEBENBURGER-09} for colloidal suspensions are qualitatively similar to those that we have discussed in earlier papers on bulk metallic glasses \cite{JSL-08,FL-10}. We believe that we understand those phenomena, and that they are not especially sensitive to the barrier-height distributions.  However, we will need to invoke nonlinear mechanisms to understand the sample-preparation methods used by Purnomo et al \cite{PURNOMO-07}.

Finally, we have considered only systems in which the ambient thermal noise is non-negligible. We have not yet extended these ideas to strictly athermal systems such as foams or granular materials. It will be interesting to do so.

\begin{acknowledgments}

We thank M. Siebenburger for sending us the data shown in Fig.\ref{G-Colloid}, and M. Falk for especially useful comments on an earlier draft of this paper. JSL was supported in part by the Division of Materials Science and Engineering, Office of Basic Energy Sciences, Department of Energy, DE-AC05-00OR-22725, through a subcontract from Oak Ridge National Laboratory. He benefitted from discussions with M. Cates, P. Sollich, M. Fuchs, D. Durian, D. Rodney, and C. Schuh, among  others at the program on glass physics at the Kavli Institute for Theoretical Physics in the spring of 2010. EB was partially supported by the Harold Perlman Family Foundation and by a grant from the Robert Rees Applied Research Fund.

\end{acknowledgments}

\end{document}